# Thermal Expansion Coefficient and Phonon Dynamics in Coexisting Allotropes of Monolayer WS₂ Probed by Raman Scattering

Deepu Kumar[1#], Birender Singh[1], Pawan Kumar[2], Viswanath Balakrishnan[2] and Pradeep Kumar[1*]

**1**School of Basic Sciences, Indian Institute of Technology Mandi, 175005, India

**2**School of Engineering, Indian Institute of Technology Mandi, 175005, India.

## Abstract:

We report a comprehensive temperature dependent Raman measurements on three different phases of monolayer WS₂ from 4K to 330K in a wide spectral range. Our studies revels the anomalous nature of the first as well as the higher order combination modes reflected in the disappearance of the few modes and anomalous temperature evaluation of the phonon self-energy parameters attributed to the detuning of resonance condition and development of strain due to thermal expansion mismatch with the underlying substrate. Our detailed temperature dependence studies also decipher the ambiguity about assignment of the two modes in literature near ~ 297 cm⁻¹ and 325 cm⁻¹. Mode near 297 cm⁻¹ is assigned as $E_{1g}$ first order Raman mode, which is forbidden in the backscattering geometry and 325 cm⁻¹ is assigned to the combination of $E_{2g}^2$ and $LA$ mode. We also estimated thermal expansion coefficient by systematically disentangling the substrate effect in the temperature range of 4K to 330K and probed its temperature dependence in 1H, 1T and 1T' phases.

#E-mail: deepu7727@gmail.com

* E-mail: pkumar@iitmandi.ac.in

# 1. INTRODUCTION

Graphene, a two dimensional form of graphite which crystalizes in the hexagonal honeycomb lattice has potential in the electronics and optoelectronic field because of very high thermal and electrical conductivity. Inspite of having excellent physical properties, it is not viable to be used for application in semiconductor industry because of zero band gap [1-2]. Owing to zero gap bottle neck in graphene, scientific community moved towards a new class of two-dimensional (2D) materials with comparable physical properties as that of Graphene but with finite band gap and therefore suitable for the semiconductor industry for next generation electronic applications. Recently, this 2D family of nanomaterials is joined by a large number of 2D transition metal dichalcogenides (TMDCs) such as $MX_2$ ($M = Mo$, W and $X = Se$, S) [3-12] with tunable indirect band gap in bulk form to the direct gap in single layer making them suitable for optoelectronics applications. In addition to the zero band gap in Graphene, the other issue in implementing in nano-devices is the contact resistance and control at the naoscale level [13-16]. The engineering of different polymorphs of monolayers of TMDC may be effectively used to overcome these barriers for their potential applications. Different polymorphs of TMDC have different physical properties, therefore coexistence of these different polymorphs in a single flake may hold the key to design future electronic devices. In particular, the $WS_2$ monolayer exists in different polymorphs, namely 2H/1H, 3R, IT and 1T', with different properties such as 2H/1H and 3R are semiconducting in nature; on the other hand 1T is metallic and 1T' is semi-metallic [15-16]. The use of these 2D materials in high density, high speed integrated electronic devices or as matrix reinforcement would also require the knowledge of the thermal expansion coefficient (TEC), in particular its dependence on the temperature as TEC is also linked with the stress in the material and indirectly may modulate the electronic properties. As in most of the application, these 2D materials are supported on some substrate, such as Si/SiO$_2$, and as results exact knowledge of the TEC become more important as a function of temperature. However, the self-heating of the electronic devices may change the electron-phonon interaction, which may be reflected in the form of reduced performance of the devices. As the mobility of charge carriers is also associated with the electron-phonon interaction, therefore any change in the strength of electron-phonon interaction will impact the performance of nano-electronic devices.

Understanding the dynamics of low energy quasi-particle excitations, phonons in particular, in 2D TMDC has played a momentous role in our current understanding of these materials [17-18]. Raman spectroscopy has been proved to be a very powerful tool to probe the quasi-particle excitations in nano materials revealing the underlying physics responsible for their excellent physical properties. In case of monolayer WS$_2$, focus in the literature has been only on 2H/1H phase [4, 19-22], as 1T and 1T' phases so far were difficult to stabilize, and even for 2H/1H phase the focus is further confined to probing only the first-order Raman active modes i.e. out of plane ($A_{1g}$) and in plane ($E_{2g}^1$) modes only. The role of multi-phonon and modes away from the Brillouin zone center have been largely unexplored. Hence, it become very important to investigate the detailed phonon dynamic of these different polymorphs of WS$_2$ monolayer. Temperature dependent phonon structure are linked to the lattice parameter and hence to the TEC. In particular, the mismatch with the underlying substrate, SiO$_2$/Si in the present case, may give rise to the strain often manifested by corrugation, buckling and as a result may modify the measured TEC; and mismatch may be more with varying the temperature. Therefore, a detailed temperature dependent Raman study focusing on all the modes is required to understand the nature of phonon dynamics and TEC of WS$_2$ monolayer as a function of temperature.

Here, we have undertaken such a study and report an estimation of the TEC of different phases of monolayer of $WS_2$ in the temperature range of 4 K – 330 K by carefully excluding the substrate effect and using the shift in the phonon mode frequency as a function of temperature. We did temperature dependent measurements from 4 K to 330 K with a very short temperature interval, allowing us to accurately decipher the temperature evolution of the phonon modes reflected in the disappearance of couple of modes and anomalous behavior of first-order as well as higher order modes observed as high as $4^{th}$ order modes. In a normal Raman scattering process, momentum conservation constrained only the observation of first-order Raman active phonon modes near zone center. However due to defects, resonance condition, this momentum conservation constrained may be violated and results into observation of number of intense modes including forbidden phonons other than the first-order modes. In the present work, we have used 532 nm (2.33 eV) wavelength, which is close to the B exciton gap in monolayer $WS_2$ and as a result of resonance effect large number of phonon modes are observed along with the first-order Raman active modes.

## 2. Experimental Details

1H, 1T and 1T' polymorphs of mono layer $WS_2$ were prepared and characterized as described in ref. [23]. Unpolarised temperature dependent Raman measurements were carried out with Horiba Labram HR evolution Raman Spectrometer in back scattering geometry. Laser power at the sample was kept very low ≤ 0.2 mW to avoid any heating effect on the sample. Scattered light was detected by using 600 grating/mm coupled with Peltier cooled Charge Coupled Device (CCD) detector. The temperature variation was carried out using closed cycle refrigerator (Montana) in wide temperature range 4-330 K, with temperature accuracy of ± 0.1 K.

## 3. Results and Discussions

### 3.0. Thermal expansion coefficient

Temperature dependence of a phonon mode in WS$_2$ may arise from two factors associated with the change in the temperature i.e. (i) thermal expansion of the lattice ($\Delta\omega_E(T)$) (ii) anharmonic effect ($\Delta\omega_A(T)$), which also result into the change in self-energy of a phonon modes. The change in the phonon mode frequency considering the above effect may be given as

$$\Delta\omega(T) = \Delta\omega_E(T) + \Delta\omega_A(T) \hspace{3cm} \text{------ (1)}$$

We note that in the present case WS$_2$ monolayer is not free standing but on the SiO$_2$/Si substrate. SiO$_2$ has negative (positive) TEC at low (high) temperature [24]. Bulk WS$_2$ is theoretical predicted to have negative TEC [25], therefore TEC mismatch with the substrate is expected to induce the strain in the WS$_2$ monolayer as the temperature is varied from room temperature. At low temperature WS$_2$ monolayer may buckle or form ripple/wrinkle due to strain caused by TEC mismatch. From Fig. 1 ( $A_{1g}$ as well $E^1_{2g}$ mode for 1T and 1H phase ) we do observed that around ~ 160K frequency of the $E^1_{2g}$ mode shows a kink and then becomes nearly constant till ~ 40K and with further decreasing the temperature it shows an upward/downward trend. Kink in the mode frequency near ~ 160K suggest the finite buckling due to strain arising from TEC mismatch. However, from 160K to 330K there is gradual decrease in the mode frequency with increasing temperature suggesting that strain from TEC mismatch is coherent in this temperature range.  As the temperature varies, thermal strain arising from TEC mismatch with the substrate should be taken into account to understand the dynamics. Therefore, change in the frequency should also be considered due to strain effect ($\Delta\omega_S(T)$), and the net change is given as [26-28]

$$\Delta\omega(T) = \Delta\omega_E(T) + \Delta\omega_A(T) + \Delta\omega_S(T) \hspace{2cm} \text{------ (2)}$$

Where the last term result from strain ($\varepsilon(T)$) due to TEC mismatch. It is given as

$$\Delta\omega_S(T) = \beta\varepsilon(T) = \beta\int_{T_0}^{T}[\alpha_{SiO_2}(T) - \alpha_{WS_2}(T)]dT$$ , where $\beta$ is the strain coefficient of a particular phonon mode, $\alpha_{SiO_2}(T)$ and $\alpha_{WS_2}(T)$ are the temperature dependent TEC of $SiO_2$ and $WS_2$ monolayer, respectively. For estimating $\alpha_{WS_2}(T)$, we have used the value of $\beta(=\frac{\partial\omega}{\partial\varepsilon})$ for $WS_2$ monolayer as $\beta_{A_{1g}} = -0.6\ cm^{-1}/\%$ and $\beta_{E_{2g}^1} = -2\ cm^{-1}/\%$ [29]. First term in eq$^n$. 2 is given as [28]

$$\Delta\omega_E(T) = \omega_0\exp[-3\gamma\int_{T_0}^{T}\alpha_{WS_2}(T)dT] - \omega_0$$ , where $\gamma$ is the Gruneisen parameter. We have taken the constant value (~ 2 for $A_{1g}$ and ~1 for $E_{2g}^1$) of the Gruneisen parameter for the purpose of estimating the TEC of $WS_2$ [25]. Second term, $\Delta\omega_A(T)$, has its origin in the change in the phonon self-energy due to anharmonic effect and is given as [27] $\Delta\omega_A(T) = A(1 + \frac{2}{e^x - 1})$ , where A is a constant parameter and $x = \hbar\omega/2K_BT$. To extract the TEC of $WS_2$ monolayer, we have fitted $A_{1g}$ and $E_{2g}^1$ mode frequency in the temperature range of 160K-330K, 4K-330K, for 1T,1H/1T' phase, respectively, using the eq$^n$. 1 above. TEC of $SiO_2$ was taken from [24] and was integrated out while estimating TEC for $WS_2$ monolayer. Figure 1 shows the estimated TEC of $WS_2$ monolayer for all three phases in the temperature range of 160K - 330K for 1T and 1H phase and 4K to 330K for 1T' phase. Our estimated TEC of $WS_2$ by systematically disentangling the substrate effect shows strong temperature dependence and its value at 300K are listed in Table-I. Our results shows, also reflected in the temperature dependence of the phonon self-energy parameters described below, that at low temperature there is significant buckling of the sample. If buckling occurs as inferred from our data then transport properties may get renormalized, also the inhomogeneous strain

coming from TEC mismatch may also results into pseudo magnetic field which will further modify the transport properties.

### 3.1 Mode of vibrations for 1H , 1T and 1T' phase in monolayer WS$_2$

Bulk WS$_2$ belongs to the point group $D_{6h}^4$ (space group - $P6_3/mmc$, # 194) with 6 atoms per unit cell resulting into 18 normal modes with the irreducible representation given a $A_{1g} + 2A_{2u} + 2B_{2g} + B_{1u} + E_{1g} + 2E_{1u} + 2E_{2g} + E_{2u}$ [30-31]. Number of Raman active modes in the bulk are four with symmetry given as $A_{1g}, E_{1g}, E_{2g}^1$ and $E_{2g}^2$. $E_{2g}^2$ is an interlayer mode hence is not observed in case of monolayer. $E_{1g}$ is an in-plane modes involving only the vibration of chalcogen atoms and is forbidden in the backscattering geometry. The first order optical phonon mode $E_{2g}^1$ is associated with the in plane vibrations of Tungsten and Sulphur atoms in opposite directions and mode $A_{1g}$ is associated with the out plane vibrations of only Sulphur atoms in opposite direction, pictorially shown in Fig. 2. As oppose to the 24 symmetry operation in case of bulk, these operation reduces to 12 in case of odd or even number of layers. Mono layer of WS$_2$ belongs to the point group $D_{3h}^1$ (space group - $P\bar{6}m2$, # 187) with 3 atoms per unit cell and 9 normal modes with irreducible representation given as $2A_2'' + A_1' + 2E' + E''$ [30, 31]. We note that in case of monolayer due to different symmetry, modes notation changes as $A_{1g}$ to $A_1'$ and $E_{2g}^1$ to $E'$, but in the literature both the notation have been used for the monolayer studies [4, 19-22, 32]. Frequency difference between the $A_{1g}$ and $E_{2g}^1$ modes manifests precise information about the number of layers present in MX$_2$ type systems. At room temperature, we observe that frequency difference between $A_{1g}$ and $E_{2g}^1$ mode is 62.1, 59.7 and 60.5 for 1H, 1T and 1T' phase, respectively, indicating the monolayer nature of WS$_2$ in line with the earlier reports [4, 19-22, 32].

Figure 2 shows the room temperature Raman spectra of three different phases, i.e. 1T, 1H and 1T' of monolayer WS$_2$ in a wide spectral range of 100-800 cm$^{-1}$. The spectra is fitted with a sum of Lorentzian function in order to extract mode frequency ($\omega$) and full width at half maximum (FWHM) of the individual mode. We have labelled all the observed modes as S1 to S11 for convenience. Table-I shows the mode frequency and the corresponding symmetry assignments of the modes for three different phases, i.e. 1T, 1H and 1T', of monolayer WS$_2$ and the mode assignment is done in accordance with the earlier reports [8, 9, 20, 33]. The first-order modes $E_{2g}^1$ (S8) and $A_{1g}$ (S9) are observed at frequency 358.1 cm$^{-1}$, 355.4 cm$^{-1}$, 357.0 cm$^{-1}$; 417.8 cm$^{-1}$, 417.5 cm$^{-1}$, 417.5 cm$^{-1}$ in 1T, 1H and 1T', phases, respectively (see Table-I). In addition to these well-known first-order Raman active modes near Brillouin zone center, we also observed first-order longitudinal acoustic (LA) mode near symmetry point $M$ in the Brillouin zone in 1T, 1H and 1T' phases at 175.3 cm$^{-1}$, 174.7 cm$^{-1}$, and 175.6 cm$^{-1}$, respectively, clearly reflecting the resonance effect. In the low frequency range an additional phonon mode around ~ 147 cm$^{-1}$ (J$_1$) is observed only in the 1T metallic phase. Furthermore, overtones and combination of phonon modes are also observed at 230.6 cm$^{-1}$, ($A_{1g}(M) - LA(M)$; S3), 325 cm$^{-1}$ ($2LA(M) - E_{2g}^2$; S5), 353.3 cm$^{-1}$ ($2LA(M)$; S7), 583.5 cm$^{-1}$ ($A_{1g}(M) + LA(M)$; S10) and at 701 cm$^{-1}$ ($4LA(M)$; S11) for 1T phase, details about other two phases is given in Table-I. We note that mode S3 and S10 are linear combination of $LA$ and $A_{1g}$ mode at the M point in Brillouin zone, the estimated frequency of $A_{1g}$ mode at M point is 407 cm$^{-1}$ and it matches very well with theoretical value [8]. The observed spectrum for all three phases shows that mode $2LA(M)$ and $E_{2g}^1$ are close to each other, where mode $2LA(M)$ is dominating as compared to $E_{2g}^1$ due to double resonance characteristic of $2LA(M)$ mode in the monolayer [19-20, 22]. Intensity of a higher order mode in a Raman process varies as $g^n$, where g is the typical electron-phonon coupling and is usually much less than one and $n$ is order of the phonon, therefore as $n$ increases intensity decreases extremely fast and higher order modes are normally not observed. We note that we observed modes as high as 4$^{th}$ order ($4LA(M)$ ; S11) modes again depicted the strong resonance phenomena in these TMDC.

## 3.2 Temperature dependence of the phonon modes

Figure 3 (a and b) shows the mode frequency ($\omega$) and full width at half maximum (FWHM) as a function of temperature for 1T phase of monolayer WS$_2$. Here, we have used a simple notation J1 and S1 to S11 to represent the observed modes and the associated symmetry details is given in Table-I. Following observation can be made from our comprehensive temperature dependent Raman measurements: (i) Frequency of the modes J1, S1, S4 and S7-S11 (S8 and S9 are shown in Fig. 1) shows normal temperature dependence (i.e. decrease in mode frequency with increasing temperature) from ~ 160 K – 330 K, which may be understood within the anharmonic phonon-phonon interaction picture. Between temperature range of ~ 160K - 40K, frequency remains nearly constant down to 40K, and below 40K, a slight upward change in the frequency of the modes J1, S1, S4 and S7-S10 is observed. (ii) Linewidths of the modes J1, S1, S4, and S8 (not shown here) shows normal behavior from 330K to ~ 160 K, i.e. linewidth decreases with decreasing temperature which may also be understood within the anharmonic model. However, below ~ 160K linewidth shows anomalous behavior i.e. it start increasing with decreasing temperature for the modes J1, S4 and S8; and for S1/S11 it suddenly increases/decreases and after that show a slight decrease/increases with decreasing temperature. Line width of mode S9 (not shown here) also shows similar behavior as that of mode S1. We note that line width of these modes does not show normal behavior in the full temperature range i.e. increase in line width with increasing temperature and interestingly many of the modes show a maxima at the middle of the temperature range. This anomaly in the linewidth may be understood from the strain caused by TEC mismatch between sample and the substrate, strain may results into change in the sample morphology such as formation of wrinkles or ripples and leading to changes in linewidth. We note that similar behavior in linewidth is also reported for Mono/bi layer of $MoS_2$ [34]. The most interesting observation is the disappearance of two modes S2 (~ 194 cm$^{-1}$) and S3 (230 cm$^{-1}$) in the vicinity of ~ 180K. From symmetry assignment, it suggests that both these modes involve the $LA$ mode and with decreasing temperature it may be out of the resonance window and as result weak modes associated with $LA$ modes are not observed or becomes too weak to be observed.

Strong spin orbit coupling along with the broken inversion symmetry in monolayer $WS_2$ allow tuning of the valley degrees of freedom, and this may be employed successfully in spintronics and valleytronics devices [35-37]. In particular, second order longitudinal acoustic ($2LA$) mode play an important role in valley depolarization due to its involvement in the inter valley scattering process [38]. Owing to the resonance effect the $2LA$ (S7) mode is stronger as compared to other first order Raman active modes, i.e. $A_{1g}$ and, $E_{2g}^1$ at room temperature (see Fig. 2). However, with decreasing temperature intensity of mode $2LA$ (S7) changes drastically as compared to mode $E_{2g}^1$ (S8) (see Fig. 2 inset and Fig. 4 for quantitative result). In fact intensity of $E_{2g}^1$ and $2LA$ mode show opposite behavior with decreasing temperature below ~ 160K. As the $2LA$ mode is an overtone so its intensity as a function of temperature is expected to follow $[n+1)*[(n+2]$ functional dependence, where $n(\omega, T) = 1/[\exp(\hbar\omega/k_B T) - 1]$ is the Bose-Einstein factor [39]. An estimate using the above relation suggest that its intensity should change by a factor of 1/3 from room temperature to 4K. Solid line in Fig. 4 for S7 mode shows fitting using the above relation, we note that the fitting is only modest. The drastic decrease in the intensity of $2LA$ mode with decrease in temperature may be due to the variation of band gap in monolayer $WS_2$ with decreasing temperature and as a result it may detune the resonant condition. In fact, intensity of the mode S7 ($2LA$) (see Fig. 4), increases with increase in temperature and become maximum at near room temperature (~ 300K) and further increase in temperature, intensity decrease which also agree with the earlier report [21]. With decreasing temperature S2 and S3 mode also disappear suggesting that excitation energy is out of the resonance window and as a result weak modes which results from combination of $2LA / LA$ modes also becomes very weak in intensity and hence disappear at low temperature.

Figure 5 (a and b) shows variation of mode frequency and FWHM as a function of temperature for 1H phase. Following observation can be made: (i) most interesting observation is the disappearance of mode S2 and S3 below temperatue ~ 160 K similar to the case of 1T phase. (ii) Frequency of the modes S1 (not shown here) and S4 shows normal temperature dependence. However, frequency of the modes S5, S7, S8 (Fig. 1a), S10 and S11 (not shown here) becomes nearly constant below ~ 160 K till ~ 40 K and below 40K it decrease/increases for the modes S5, S7,S8/S10, respectively. (iii) FWHM of the mode S1and S9 (not shown here) and S5, S10 show normal behavior in the temperature range of 4 K to ~ 160 K, but above 160 K it shows anomalous behavior i.e. decreases in linewidth with increasing temperature. (iv) Linewidth of mode S4 shows anomalous behavior in temperature range from 4 K to ~ 160 K and normal above it. However, linewidth of mode S8 (not shown here) shows anomalous temperature dependence, it remains nearly constant in the temperature range of 4 K -160 K and above 160 K it start decreasing with increasing temperature. Figure 6 (a and b) shows variation of mode frequency and linewidth for 1T' phase. Following observation can be made: (i) frequency of the modes S4-S7, S8-S9 (see Fig. 1a, b) show normal temperature dependence (ii) Line-width of the mode S5 and S8 and S9 (not shown here) show normal behavior, however mode S7 shows anomalous change in the slope of the linewidth below ~ 160 K. The behavior of line-width of mode S4 is very interesting and variation in line-width with temperature can be divided into three region, line-width shows normal behavior in the temperature range 4 K to ~ 160 K, and then shows a sudden drop around this temperature and with further increase in temperature it start increasing till the highest recorded temperature (330K). Intensity of the mode $A_{1g}$ for the WS$_2$ monolayer is reported to be increasing with decreasing resonance conditions [21], i.e. if resonance condition is detuned then intensity of $A_{1g}$ mode should start increasing and that of $2LA$ mode should be decreasing fast. We note that intensity of mode $A_{1g}$, see Fig. 4 (b), in 1T and 1T' phase shows very interesting temperature dependence below ~ 160K i.e. it start increasing with further decrease in temperature, increase at 4K is as high as 400% w.r.t its value at 160K, again suggesting that resonance condition is detuning at lower temperature.

We note that in all the three phases a quite strong mode is observed near ~ 325 cm$^{-1}$ (S5), its origin is not very clear and the corresponding symmetry assignment is under debate, however it has been associated to the $E_{1g}$ mode [8, 40] which is forbidden in the backscattering geometry as well as to a combination of $2LA$ and $E_{2g}^2$ mode, i.e. $2LA(M) - E_{2g}^2(M)$ [33, 44]. To decipher this confusion in the literature on the assignment of this mode we plotted the intensity of mode S5 as a function of temperature, see Fig. 4a. We note that the change in its intensity is huge with increasing temperature, similar to other prominent overtone or combination modes. Usually, for a first order phonon mode intensity should be $\propto (1+n)$, a rough estimate of the change in intensity for the first order mode would give a change of ~25 % at 300K w.r.t its value at 4K, however a dramatic change in the intensity of mode S5, clearly suggests that the origin of this mode is not first-order $E_{1g}$ mode but rather a combination mode. We tried to fit this combination mode with the function $(1 + n_2) * n_1$, which describe the temperature variation of the intensity of a mode which arise from the difference of two modes [39] and found that the intensity fit is modest again suggesting that the mode near 325 cm$^{-1}$ is not a first order mode as suggested in the literature rather a combination mode.

Another interesting observation is the mode S4, we note that in literature [8, 9, 20, 40, and 44] it has been attributed to the combination mode i.e. $2LA(M) - 2E_{2g}^2(M)$ as well as first-order $E_{1g}$ mode. Top panel in Figure 4 (a) shows the intensity variation of this mode and interestingly change in its intensity suggest this to be a first-order mode rather than an overtone or a combination mode as postulated in the earlier reports on this system. Solid line shows that the fit is very good, fitted with the first-order mode functional form i.e. (1+n). Based on our experimental observation we attribute this mode to the first order $E_{1g}$ mode which is forbidden in backscattering. Observation of backscattering forbidden phonons in various 2D TMDC was reported due to the resonant Raman scattering effect [40, 41]. Here, we also attribute this mode to the resonant Raman effect as well. Under resonance condition, the observation of backscattering forbidden phonons may be understood via Fröhlich mechanism of coupling between exciton and phonons [42]. Under resonance condition, exciton-phonon interaction results into finite intensity of a forbidden phonon mode is $\propto (aq)^2$, where $a$ is the characteristic length scale of the excited state i.e. Bohr radius of the exciton and $q$ is the wave vector of the phonon (i.e. $\vec{q} = \vec{k}_i - \vec{k}_f$) [42]. Therefore, when $a$ is large as compared to the lattice constant, forbidden phonons may be observed under resonance condition. These conditions may be easily satisfied for the case for WS$_2$ as exciton in these systems have very large radius [43] and as a results they can couple to the phonons resulting into finite intensity even for the forbidden phonons.

## CONCLUSION

In summary, in this work we performed a detailed Raman studies on three different phases of monolayer $WS_2$ i.e.1H, 1T and 1T' in a broad spectral and wide temperature range. We assigned the mode near 297 cm$^{-1}$ and 325 cm$^{-1}$ as first order mode with $E_{1g}$ symmetry which is forbidden in back scattering and a combination mode, respectively; deciphering the uncertainty on the assignment of the  these modes in the literature. Our report is a first comprehensive temperature dependence studies covering first order as well as overtone /combination modes. Some of the combination modes disappear in lower temperature range attributed to detuning of the resonance condition reflected in the anomalous intensity evaluation of the $LA$ mode as well as $A_{1g}$ with lowering temperature. Our detailed analysis also revealed the strong temperature dependence of the bare (excluding substrate effect) thermal expansion coefficient of $WS_2$ for all the three phases.

## Acknowledgements

PK acknowledge DST Nano Mission, India for the financial support and AMRC, IIT Mandi, for the experimental facilities.

**Table-I:** Room temperature thermal expansion coefficients for 1T, 1H and 1T' phase of monolayer WS$_2$ extracted using $E^1_{2g}$ and $A_{1g}$ modes.

| WS$_2$ Phase | $\alpha(E^1_{2g})$ $10^{-5}$ K$^{-1}$ | $\alpha(A_{1g})$ $10^{-5}$ K$^{-1}$ |
|---|---|---|
| 1T | 1.431 | 0.716 |
| 1H | 1.403 | 0.074 |
| 1T' | -0.238 | 0.208 |

**Table-II:** List of the observed modes, along with their symmetry assignments, frequency at room temperature for 1T, 1H and 1T' phase of monolayer WS$_2$. Units are in cm$^{-1}$.

| Mode Assignment | Frequency $\omega$ | | |
|---|---|---|---|
| | 1T | 1H | 1T' |
| **J1** | $147.2 \pm 2.1$ | | |
| **S1** [ $LA(M)$ ] | $175.3 \pm 0.5$ | $174.7 \pm 1.4$ | $175.7 \pm 0.8$ |
| **S2** [ $LA(K)$ ] | $193.9 \pm 2.3$ | $194.8 \pm 3.1$ | $193.2 \pm 2.4$ |
| **S3** [ $A_{1g}(M) - LA(M)$ ] | $230.6 \pm 1.7$ | $231.5 \pm 2.8$ | $231.5 \pm 2.3$ |
| **S4** [ $E_{1g}(\Gamma)$ ] | $296.0 \pm 0.7$ | $296.1 \pm 1.0$ | $296.9 \pm 0.8$ |
| **S5** [ $2LA(M) - E^2_{2g}$ ] | $324.0 \pm 0.4$ | $324.4 \pm 0.7$ | $325.7 \pm 0.5$ |
| **S6** | $345.0 \pm 0.8$ | $342.7 \pm 1.5$ | $345.8 \pm 0.8$ |
| **S7** [ $2LA(M)$ ] | $353.3 \pm 0.1$ | $350.1 \pm 0.5$ | $352.1 \pm 0.4$ |
| **S8** [ $E^1_{2g}(\Gamma)$ ] | $358.1 \pm 0.1$ | $355.4 \pm 0.5$ | $357.0 \pm 0.4$ |
| **S9** [ $A_{1g}(\Gamma)$ ] | $417.8 \pm 0.4$ | $417.5 \pm 0.6$ | $417.6 \pm 0.8$ |
| **S10** [ $A_{1g}(M) + LA(M)$ ] | $583.5 \pm 1.6$ | $581.2 \pm 2.3$ | $582.7 \pm 1.7$ |
| **S11** [ $4LA(M)$ ] | $701.8 \pm 1.3$ | $699.5 \pm 2.9$ | $703.5 \pm 1.9$ |

**FIGURE CAPTION:**

**FIGURE 1:** (Color online) Temperature dependence of the phonon frequencies of modes (a) $E_{2g}^1$ (S8) and (b) $A_{1g}$(S9) for 1T, 1H and 1T' phase of monolayer WS$_2$. Solid red line are the fitted curve as described in the text and solid blue lines are guide to the eye. (c, d) Thermal expansion coefficient corresponding to the $E_{2g}^1$ (S8) and $A_{1g}$ (S9) modes for 1T, 1H and 1T' phase in the temperature range of 160K/4K-330K, respectively.

**FIGURE 2:** (Color online) Unpolarized Raman spectra of 1T, 1H and 1T' phase of monolayer WS$_2$ recorded at room temperature. The thick solid red line show the total sum of Lorentizian fit and thin blue line show individual fit. Inset figures show temperature dependence of the Raman spectra for three distinct phases in the spectral range 290- 440 cm$^{-1}$ showing the evolution of mode S7-S9.

**FIGURE 3:** (Color online) Temperature dependence of the phonon frequencies and full width at half maxima (FWHM) of modes (a) J1, S1 and S4; (b) S7, S10 and S11 for 1T phase. Solid blue lines are guide to the eye. Shaded part depicts the region where mode frequency show anomalous behavior.

**FIGURE 4:** (Color online) Temperature dependence of the normalized intensity of phonon modes (a) S4, S5, S7 and S8 for 1T phase. Solid red line are the fitted curve as described in the text. (b) Temperature dependence of the normalized intensity of phonon mode S9 for 1T, 1H and 1T' Phase.

**FIGURE 5:** (Color online) Temperature dependence of the phonon frequencies and full width at half maxima (FWHM) of modes S4, S5, S7 and S10 for 1H phase. Solid blue lines are guide to the eye. Shaded part depicts the region where mode frequency show anomalous behavior.

**FIGURE 6:** (Color online) Temperature dependence of the phonon frequencies and full width at half maxima (FWHM) of modes S4, S5, S7 and S8 for 1T' phase. Solid blue lines are guide to the eye.

**FIGURE 1:**

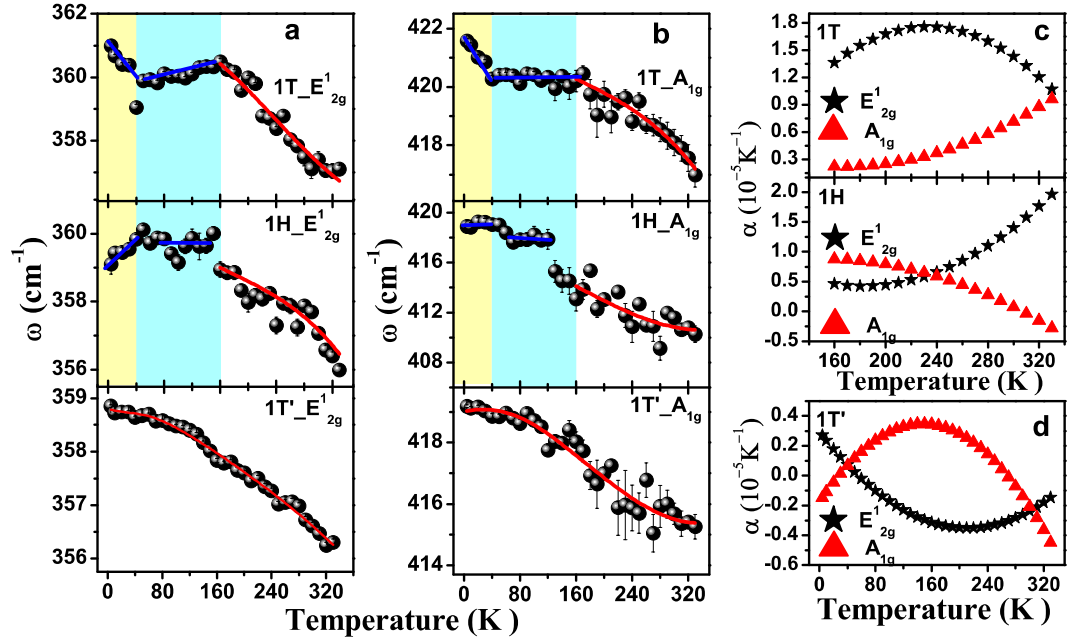

**FIGURE 2:**

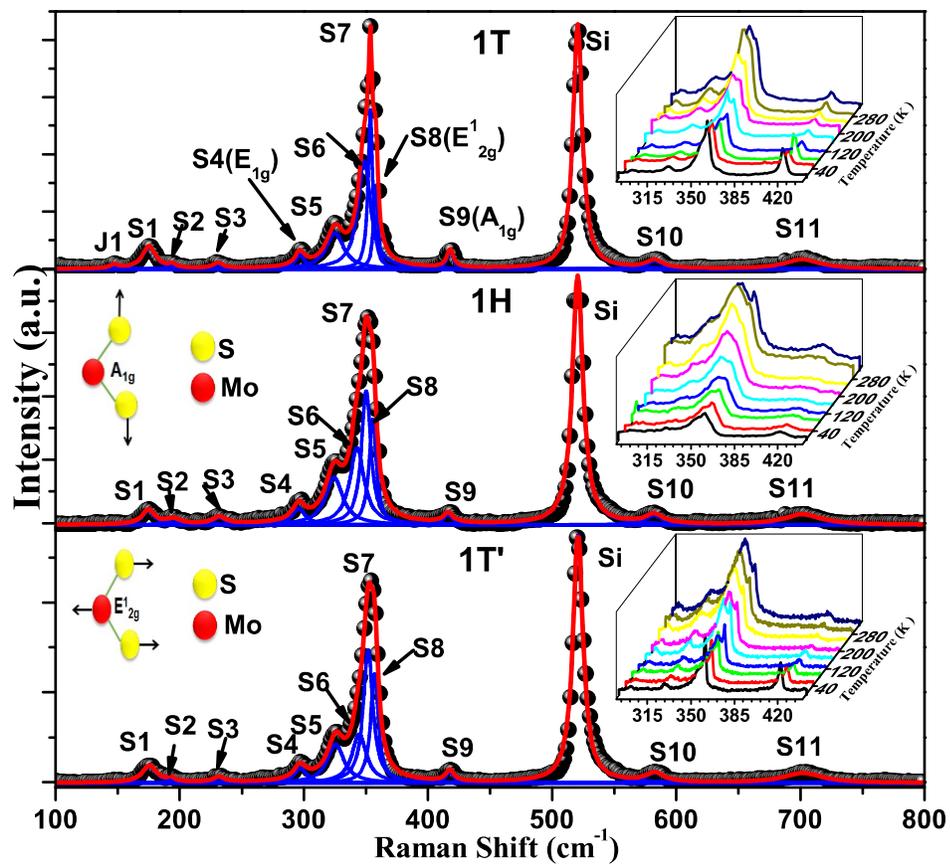

**FIGURE 3:**

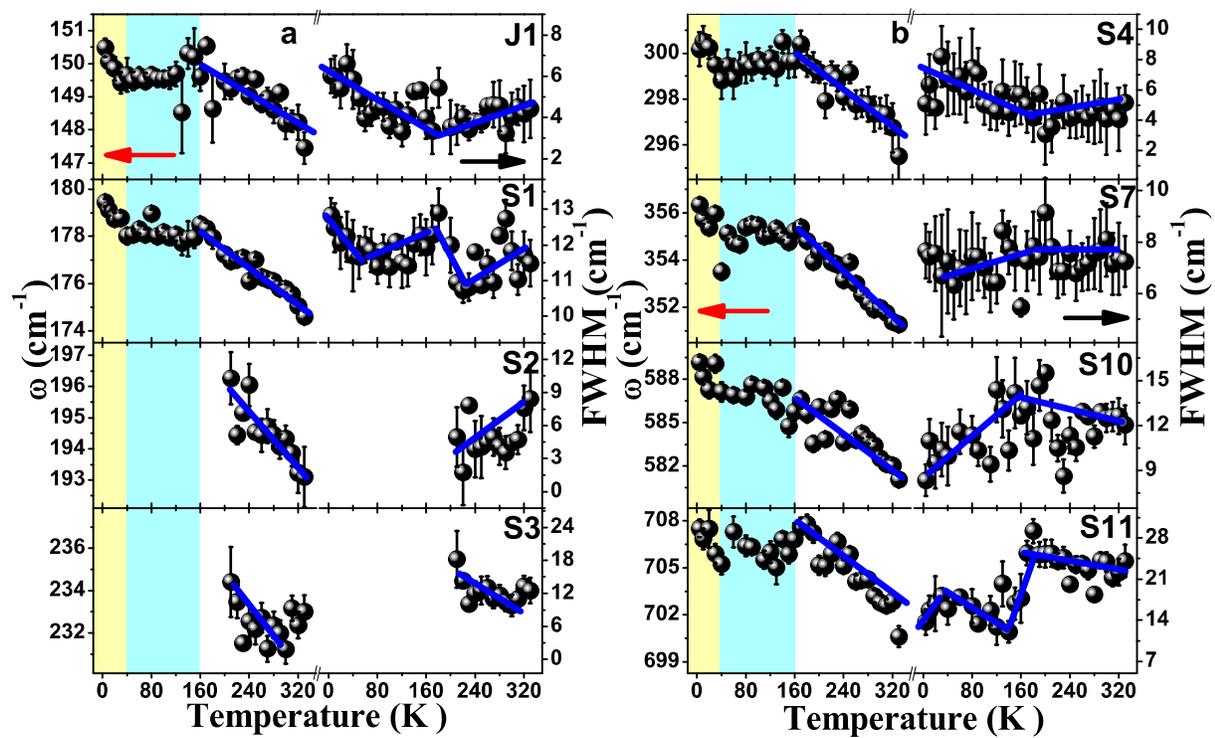

**FIGURE 4:**

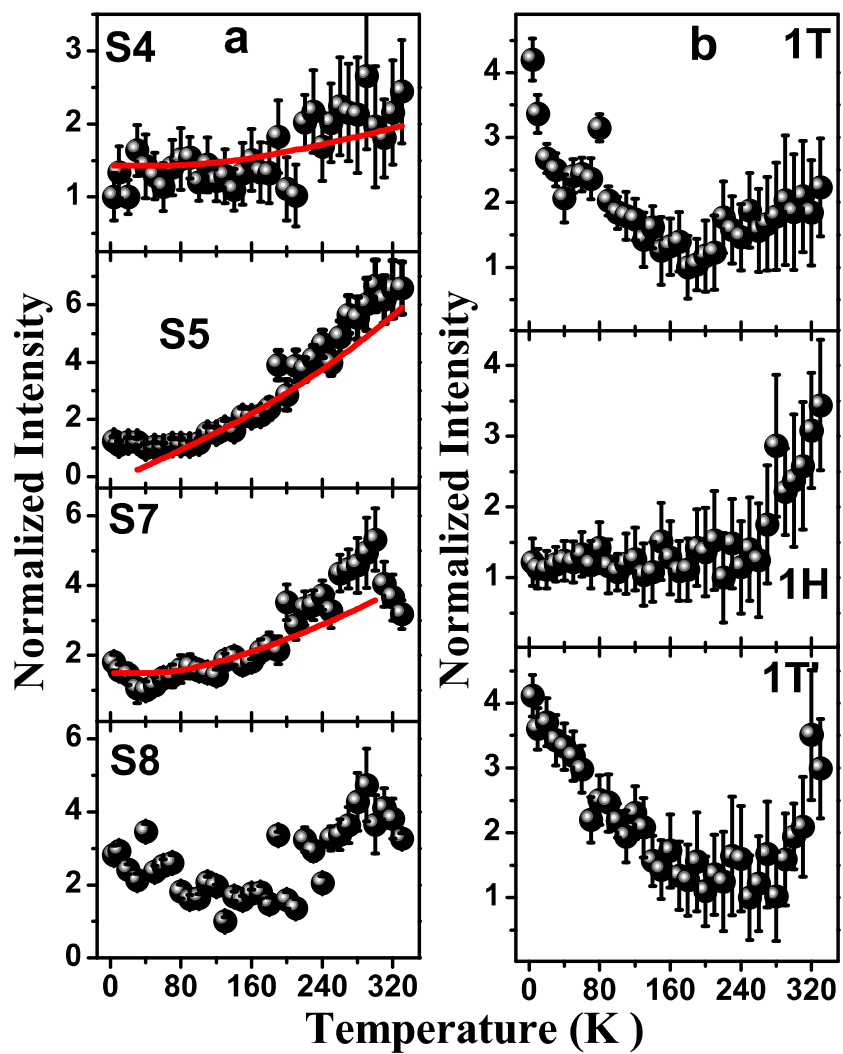

**FIGURE 5:**

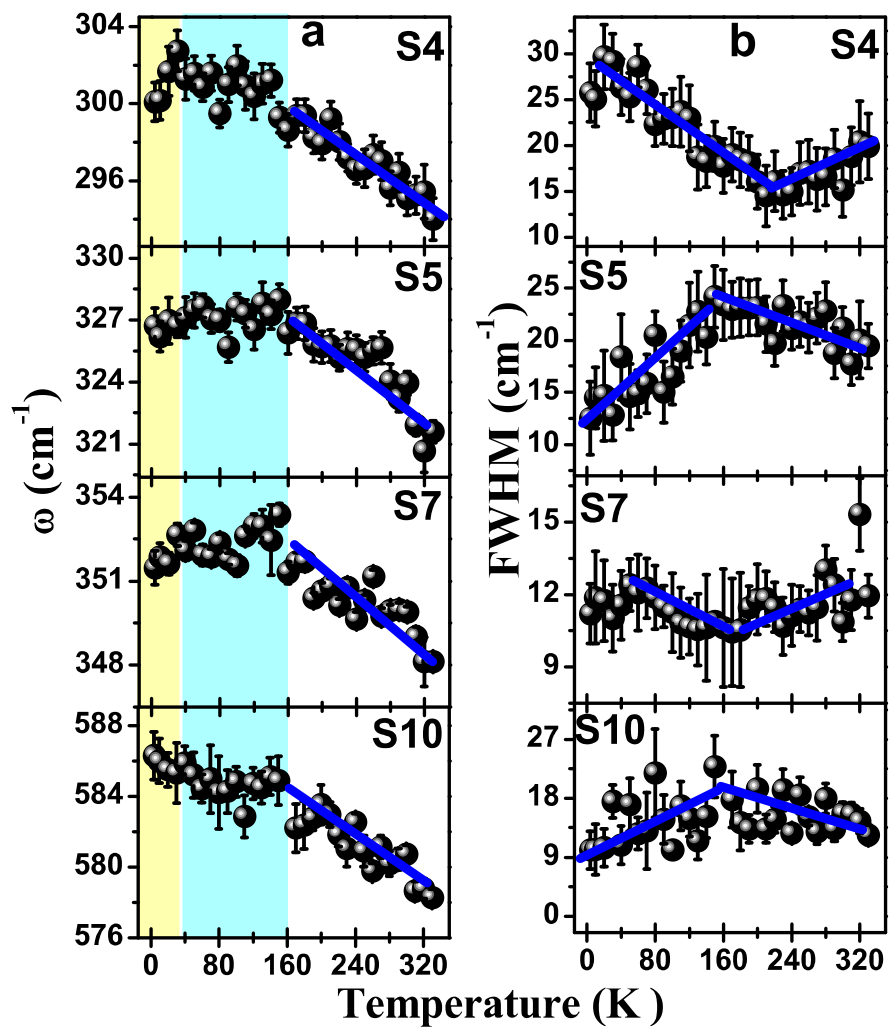

**FIGURE 6:**

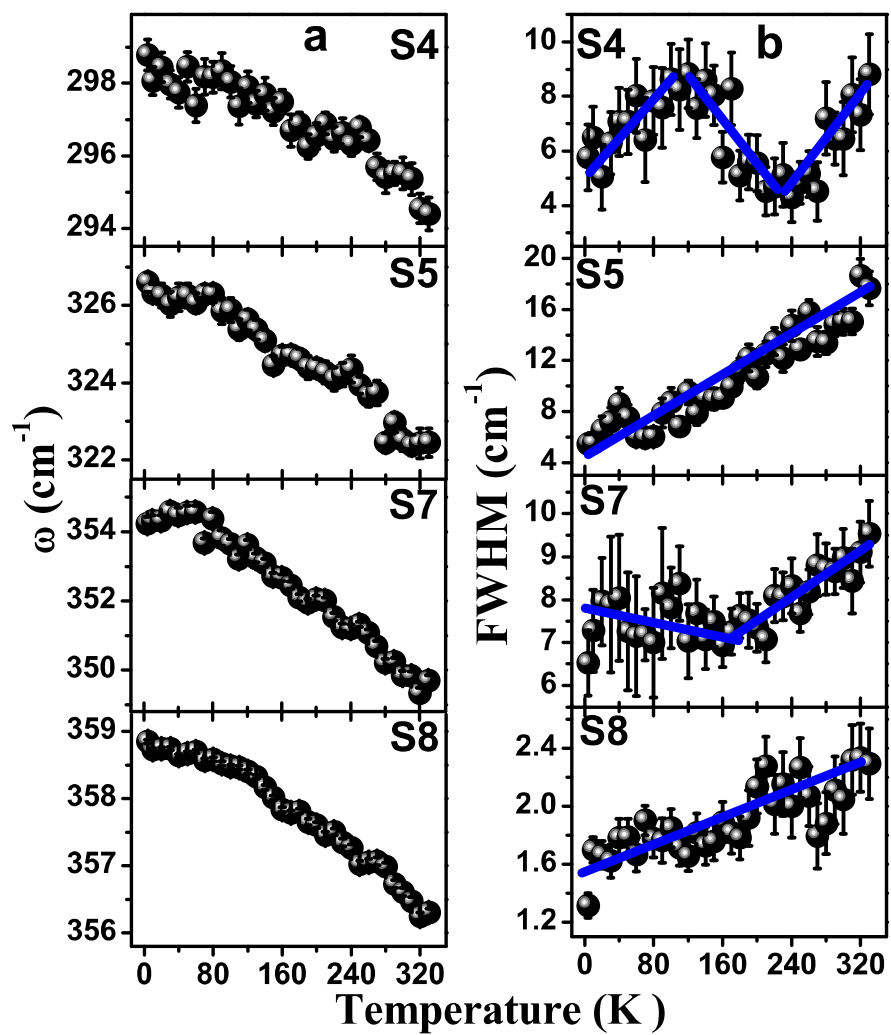